\documentclass[%
superscriptaddress,
showpacs,preprintnumbers,
amsmath,amssymb,
aps,
prl,
twocolumn
]{revtex4-1}
\usepackage{ulem}
 \usepackage{graphicx}
\usepackage{dcolumn}
\usepackage{bm}
\usepackage{color}
\usepackage{threeparttable}

\begin {document}

\title{
TDP-43 multidomains and RNA modulate interactions and viscoelasticity in biomolecular condensates
}

\author{Yui Matsushita}
\affiliation{Department of System Design Engineering, Keio University, Yokohama, Kanagawa 223-8522, Japan}

\author{Ikki Yasuda}
\affiliation{Department of Mechanical Engineering, Keio University, Yokohama, Kanagawa 223-8522, Japan}

\author{Fuga Watanabe}
\affiliation{Department of System Design Engineering, Keio University, Yokohama, Kanagawa 223-8522, Japan}

\author{Eiji~Yamamoto}
\email{eiji.yamamoto@sd.keio.ac.jp}
\affiliation{Department of System Design Engineering, Keio University, Yokohama, Kanagawa 223-8522, Japan}




\begin{abstract}
RNA-binding proteins form biomolecular condensates with RNA through phase separation, playing crucial roles in various cellular processes.
While intrinsically disordered regions (IDRs) are key drivers of phase separation, additional factors such as folded domains and RNA also influence condensate formation and physical properties.
However, the molecular mechanisms underlying this regulation remain elusive.
Here, using molecular dynamics simulations, we investigate how the multidomain structure of TDP-43, which consists of its IDR, RNA recognition motifs (RRMs), and N-terminal domain (NTD), interacts with RNA and affects the characteristics of phase separation.
Our analysis reveals that interaction sites within the IDR undergo dynamic rearrangement, driven by key residues that depend on the specific combination of folded domains.
Upon RNA binding, several intermolecular interactions of TDP-43 are replaced by TDP-43-polyA interactions, altering viscoelastic properties of the condensate.
Specifically, RRMs enhance viscosity, whereas the NTD reduces it.
The presence of polyA increases elasticity, making viscosity and elasticity comparable in magnitude.
These findings suggest that the multidomain structure of TDP-43 and its RNA interactions orchestrate condensate organization, modulating their viscoelastic properties.
\end{abstract}

\maketitle

\section*{Introduction}
Ribonucleoprotein bodies are biomolecular condensates formed via phase separation of proteins and nucleic acids.
They play key roles in various cellular processes such as transcription, signal transduction, and stress responses~\cite{ShinBrangwynne2017, FayAnderson2018, youn2019properties, wiedner2021s, wadsworth2024rna}.
Multivalent interactions are key molecular mechanisms driving phase separation, mediated by intrinsically disordered regions (IDRs) and folded domains of intrinsically disordered proteins (IDPs), as well as surrounding nucleic acids~\cite{banani2017biomolecular, boeynaems2018protein}.
IDRs facilitate non-specific intermolecular interactions, such as electrostatic forces, $\pi$-$\pi$ stacking, $\pi$-cation interactions, and hydrophobic interactions~\cite{holehouse2024molecular}.
Folded domains engage in both specific and non-specific interactions within condensates, modulating phase separation propensity~\cite{mohanty2022principles,dorner2024tag}.
In addition, nucleic acids further influence phase separation through electrostatic interactions via their negatively charged backbones.

RNA recognition motifs (RRMs) in RNA-binding proteins specifically recognize RNA or DNA sequences~\cite{li2012phase, zhang2015rna}. 
Proteins composed of multiple concatenated RRMs from various RNA-binding proteins exhibit phase separation driven by specific RRM-RNA interactions~\cite{lin2015formation}.
Studies on hnRNPA1 have shown that charged patches on RRMs interact with oppositely charged regions of IDRs, promoting phase separation~\cite{martin2021interplay}.
Additionally, fusing a hydrophilic tag to hnRNPA1 increases solubility and inhibits phase separation, suggesting that the surface properties of folded domains regulate condensate formation~\cite{martin2021interplay}.

In addition to its role in sequence-specific binding, RNA exhibits a biphasic effect on phase separation: it promotes phase separation at low concentrations, whereas at high concentrations, it inhibits phase separation~\cite{molliex2015phase, maharana2018rna}.
This phenomenon can be partially explained by reentrant phase behavior driven by electrostatic interactions~\cite{banerjee2017reentrant}.
RNA sequences are further characterized by their ability to form $\pi$ interactions.
In particular, adenine bases, which have a stronger tendency to engage in $\pi$ interactions with arginine residues, exhibit a higher propensity for phase separation~\cite{alshareedah2019interplay}.
In contrast, while RNA does not affect the phase separation propensity of LAF-1, it modulates viscoelastic properties of condensates~\cite{elbaum2015disordered}.
Viscoelasticity is tuned by the underlying molecular interactions~\cite{jawerth2020protein, alshareedah2021, alshareedah2024}.
Therefore, phase separation is governed by the interplay between IDRs, folded domains, and nucleic acids, with their effects varying depending on protein properties and molecular composition.

TAR-DNA-binding protein 43 (TDP-43) is an RNA-binding protein primarily localized in the nucleus and, under stress conditions, in the cytoplasm~\cite{fox2018paraspeckles, marcelo2021stress, lang2024tdp}.
TDP-43 can form pathological amyloid fibrils, which are a characteristic feature of amyotrophic lateral sclerosis and frontotemporal lobar degeneration~\cite{tziortzouda2021triad}.
TDP-43 consists of four domains: a disordered C-terminal domain (CTD), two RRMs (RRM1 and RRM2), and a folded N-terminal domain (NTD) (Fig.~\ref{fig1}A).
The CTD is rich in Gly, Ser, Asn, Gln, and contains aromatic residues~\cite{udan2011implications, loughlin2019tdp, chien2021different}.
In particular, CTD alone undergoes phase separation both in \textit{vivo} and \textit{vitro}, similar to the full-length protein~\cite{tziortzouda2021triad}. 
In solution, the CTD shows transient $\alpha$-helical structures at residues 320-340 , and mutations in residues or other hydrophobic residues disrupt phase separation~\cite{conicella2016mutations, conicella2020tdp, li2018tar}.
Moreover, while RRMs specifically bind to UG or TG repeats in RNA or DNA~\cite{lukavsky2013molecular}, the NTD tends to oligomerize, potentially facilitating phase separation~\cite{afroz2017functional, carter2021n}.
TDP-43 undergoes phase separation in conjunction with RNA~\cite{babinchak2019role,loganathan2020or}, leading to RNA-induced inhibition of amyloid formation.
This inhibition mechanism is based on both specific binding of RNA to RRM~\cite{huang2013inhibition,mann2019rna,grese2021specific}, and sequence-independent RNA effects~\cite{maharana2018rna}.

Molecular dynamics (MD) simulations have been used to elucidate the molecular mechanisms underlying the phase separation of TDP-43.
All-atom (AT) MD simulations have investigated the $\alpha$-helix propensity of the CTD in solution~\cite{conicella2020tdp}, the tandem arrangements of RRMs~\cite{ozguney2024rna}, the single-chain conformation in solution~\cite{mohanty2023synergy}, and the dimerization of the CTD~\cite{tang2024uncovering}.
To simulate large systems containing multiple proteins over longer timescales, coarse-grained (CG) models such as the hydropathy scale (HPS) model, which represents each amino acid or nucleotide as a single bead~\cite{dignon2018sequence, regy2020sequence, tesei2023improved, regy2021improved, joseph2021physics}, have been used.
While CG models lack atomistic-level resolution, these enable the investigation of phase separation dynamics that are inaccessible to AT models.
Simulations of full-length TDP-43 condensation using the HPS model demonstrated that the presence of RNA modulates the viscosity of the condensates~\cite{tejedor2021rna}.
Combined AT and CG MD simulations have suggested that various inter-domain in full-length TDP-43 contribute to phase separation ~\cite{mohanty2023synergy, ingolfsson2023multiscale, mohanty2024complex}.
Despite the molecular insights provided by these computational studies, an integrative understanding of the cooperative interactions between TDP-43 domains and their interplay with RNA remains enigmatic, as does the effect of these interactions on viscoelastic properties of TDP-43 condensates.

The amino acid composition of TDP-43 varies among its domains, which is thought to underlie their distinct roles in phase separation and RNA interactions ({Fig.~S1}).
Specifically, the CTD is intrinsically disordered and enriched in polar residues, the RRMs are abundant in charged residues, and the NTD contains a high proportion of acidic residues.
In this study, we investigate how these domains contribute to the formation and physical properties of biomolecular condensates using CG MD simulations.
To systematically analyze the effects of folded domains and RNA, we compare four TDP-43 constructs containing different combinations of folded domains (CTD, CTD with RRM2, CTD with both RRM1 and RRM2, and the full-length protein), both in the presence and absence of RNA.
Our analysis reveals that folded domains and RNA induce dynamic rearrangements of molecular interactions, leading to distinct viscoelastic properties in the condensates.

\section*{Methods}
\subsection*{Coarse-Grained Model of TDP-43 and RNA}
TDP-43 consists of three folded domains (NTD, RRM1, RRM2) and one IDR at the C terminal (Fig.~\ref{fig1}A).
We used the one-bead-per-residue model, Mpipi~\cite{joseph2021physics}, as the force field parameter of TDP-43.
The initial structure of TDP-43 was built from an Alphafold-predicted structure (AlphaFold DB: AF-Q13148-F1~\cite{varadi2024alphafold}) where CG beads were placed at the positions of the $\rm{C}_{\alpha}$ atoms.
The bonded potential between neighboring residues was modeled by a harmonic potential with an equilibrium length of $0.381 \, \mathrm{nm}$ and a spring constant of $8030 \, \mathrm{kJ mol^{-1} nm^{-2}}$.
Folded structures were maintained using an elastic network~\cite{periole2009combining}, which applies a harmonic potential with a spring constant of $500 \, \mathrm{kJ mol^{-1} nm^{-2}}$, where residue pairs within $0.9 \, \mathrm{nm}$ were connected.
While residues 320--340 are suggested to form a transient helical structure~\cite{conicella2016mutations,conicella2020tdp}, we did not apply restraints to the helical structure, considering the low confidence score in the Alphafold prediction.
Short-range interactions were represented by the Wang-Frenkel potential~\cite{joseph2021physics}, with a cutoff distance of 3$\sigma$, where $\sigma$ is a parameter determined for each pair of amino acids~\cite{joseph2021physics}.
Since interactions from buried residues are overestimated~\cite{kim2008coarse}, previous studies have employed various approaches to address this, such as reducing interactions in folded-domain residues~\cite{dignon2018sequence,tejedor2021rna,joseph2021physics}, increasing the temperature in simulations~\cite{mohanty2024complex}, or shifting bead positions to the center of mass of residues~\cite{cao2024coarse}.
In this study, we reduced the interaction strength within the folded domains by 30\%~\cite{joseph2021physics}.
Electrostatic interactions were modeled using the Debye-H\"uckel potential, with a Debye length of $1.26 \, \mathrm{nm^{-1}}$ and a cutoff distance of $3.5 \, \mathrm{nm}$.

We used a 250-nucleotide polyadenine (polyA) chain as our RNA model (Fig.~\ref{fig1}A), which was modeled using the one-bead-per-nucleotide model in Mpipi~\cite{joseph2021physics}.
Bonded interactions between neighboring adenines were modeled using a harmonic potential with an equilibrium length of $0.5 \, \mathrm{nm}$ and a spring constant of $8030 \, \mathrm{kJ mol^{-1} nm^{-2}}$.
The charge of each nucleotide was set to $-0.75 e$.

\subsection*{MD Simulations of TDP-43 and RNA Condensates in Slab Systems}
We performed CG MD simulations of biomolecular condensates in slab systems with periodic boundary conditions (Fig.~\ref{fig1}B).
The compositions and system sizes of the eight simulated systems are listed in Tab.~\ref{tab:slab}.
These include four types of multidomain constructs (LCD, RRM2-LCD, RRM1-RRM2-LCD, and FULL) with and without RNA chains.
The number of TDP-43 chains, $N_{\rm{Pro}}$, was adjusted to ensure that the total number of beads was approximately the same across all eight systems.
For systems containing polyA, the number of RNA chains, $N_{\rm{RNA}}$, was set to achieve a TDP-43:polyA bead ratio of $5.7:1$, corresponding to a mass ratio of approximately $2:1$.

Simulations were performed using LAMMPS2022~\cite{THOMPSON2022108171}.
We first prepared a smaller condensate as the basic unit for constructing a larger system.
One-third of the total chains were randomly placed in a smaller box.
Biased simulations for $10 \, \mathrm{ns}$ were performed, applying a flat-bottom potential to gather proteins in a region.
Subsequently, an unbiased simulation was conducted for $50 \, \mathrm{ns}$.
The resulting smaller condensate was duplicated and shifted along the $z$-axis to create the initial structure for the production run (Tab.~\ref{tab:slab}).
The production runs were performed for $6.5 \, \mathrm{\mu s}$, and trajectories in the last $3.5 \, \mathrm{\mu s}$ were used for analysis.
In these simulations, the Langevin integrator was employed with a timestep of $10 \ \mathrm{fs}$ and a damping parameter $\tau$ of $5 \, \mathrm{ps}$, which is the inverse of the friction coefficient $\gamma=1/\tau$, at a temperature of $310 \, \mathrm{K}$.

\begin{table*}[tb]
        \caption{Systems of slab simulations.}
        \label{tab:slab}
        \centering
        \begin{threeparttable}
                \begin{tabular}{c c c c}
                        \hline
                        Domain constructs & $N_{\rm{Pro}}$ & $N_{\rm{RNA}}$ & [$L_{x},\,L_{y},\,L_{z}$]~(nm) \\\hline
                        CTD  & 288 & 30 & \\
                        RRM2-CTD & 192 & 30 & $L_{x}=L_{y}=15$\\
                        RRM1-RRM2-CTD & 138 & 30 & $L_{z}=300$ \\
                        NTD-RRM1-RRM2-CTD\ (FULL) & 102 & 30 & \\
                        \hline
                        CTD & 288  & - & \\
                        RRM2-CTD & 192 & - & $L_{x}=L_{y}=15$ \\
                        RRM1-RRM2-CTD & 138 & - & $L_{z}=600$ \\
                        NTD-RRM1-RRM2-CTD\ (FULL) & 102 & - & \\
                        \hline
                \end{tabular}
        \end{threeparttable}
\end{table*}

\subsection*{Mass Density Profile}
Along the long axis of the slab system, the mass density of the bins was calculated based on the mass and positions of individual beads, using a bin width of $1 \, \mathrm{nm}$.
Before calculating the mass density profile, the trajectories were translated so that the center of mass of the condensates was positioned at the center of the box.

\subsection*{Contact Map}
A pair of beads was considered to be in contact if the distance between them was less than the cutoff distance $r_{C}$.
The cutoff distance was defined as $r_{\mathrm{C}}=2^{1/6}\sigma$, where $\sigma$ is a parameter, specific bead pairs in the Mpipi model~\cite{joseph2021physics}.
Normalizing the number of contacts by the number of TDP-43 chains, we computed contact ratio, 
\begin{equation}
\alpha = N_C/N
\end{equation}
where $N$ is the number of TDP-43 chains, and $N_{\rm{C}}$ is the number of contacts between either TDP-43--TDP-43 or TDP-43--RNA interactions.

\begin{table*}[tb]
        \caption{Systems of cubic simulations.}
        \label{tab:cubic}
        \centering  
        \begin{threeparttable}
                \begin{tabular}{c c c c}
                        \hline
                        Domain constructs & $N_{\rm{Pro}}$ & $N_{\rm{RNA}}$ & [$L_{x},\,L_{y},\,L_{z}$]~(nm)  \\\hline
                        CTD & 389 & 40 & \\
                        RRM2-CTD & 212 & 30 & $L_{x}=L_{y}$\\
                        RRM1-RRM2-CTD & 179 & 38 & $=L_{z}=41$\\
                        NTD-RRM1-RRM2-CTD\ (FULL) & 108 & 30 &\\
                        \hline
                        CTD & 340 & - & \\
                        RRM2-CTD & 118 & - & $L_{x}=L_{y}$ \\
                        RRM1-RRM2-CTD & 279 & - & $=L_{z}=41$ \\
                        NTD-RRM1-RRM2-CTD\ (FULL) & 227 & - & \\
                        \hline
                \end{tabular}
        \end{threeparttable}
\end{table*}

\subsection*{Shear Modulus and Viscosity}
To determine the viscoelastic properties of condensates, we performed MD simulations using a cubic system with periodic boundary conditions (Tab.~\ref{tab:cubic}).
The box sizes were set to be nearly the same for all cubic system, and the numbers of $N_{\rm Pro}$ and $N_{\rm RNA}$ were adjusted to match the density of the dense phase observed in the slab systems.
Production runs were conducted for $20 \, \mu \mathrm{s}$ under the same simulation conditions as those used in the slab simulations.
Viscoelasticity was evaluated by calculating the shear modulus from the stress tensor~\cite{ramirez2010efficient, tejedor2021rna},
\begin{widetext}
\begin{eqnarray}
G(t) &=& \frac{V}{5k_{B}T} \left[
    \left\langle \sigma_{xy}(0) \sigma_{xy}(t) \right\rangle +
    \left\langle \sigma_{xz}(0) \sigma_{xz}(t) \right\rangle +
    \left\langle \sigma_{yz}(0) \sigma_{yz}(t) \right\rangle
\right] \nonumber \\
&+& \frac{V}{30k_{B}T} \left[
    \left\langle N_{xy}(0) N_{xy}(t) \right\rangle +
    \left\langle N_{xz}(0) N_{xz}(t) \right\rangle +
    \left\langle N_{yz}(0) N_{yz}(t) \right\rangle
\right] .
\end{eqnarray}
\end{widetext}
In the short-time regime of $G(t)$, intramolecular interactions dominate, corresponding to elastic behavior primarily governed by local molecular vibrations.
Conversely, in the long-time regime, intermolecular interactions dominate, leading to viscous behavior driven by large-scale molecular rearrangements and relaxation.
The relaxation time in the short-time regime is defined as $t_0$.
The viscosity can be obtained by integrating the relaxation modulus over time, $\eta = \int_{0}^{\infty} G(t) dt$.
The integral in the short-time regime was directly computed using the trapezoidal rule.
However, because the long-time regime in the actual data converges slowly, the long-time behavior was fitted to the Maxwell model to compute the integral value analytically~\cite{tejedor2021rna},
\begin{equation}
    G_{\rm{M}}(t) = \sum_{i}^N h_{i} \exp \left( -\frac{t}{\tau_{i}} \right),
     \label{eq_maxwell}
\end{equation}
where $N$ is the number of exponents of the Maxwell model, $h_i$ is the intensity, and $\tau_{i}$ is the relaxation time.
These parameters were determined by fitting the data using Reptate~\cite{boudara2020reptate}.
The viscosity was then calculated as:
\begin{equation}
\eta = \eta (t_{0}) + \int_{t_{0}}^{\infty} dt G_{\rm{M}}(t).
\label{eq_viscosity}
\end{equation}
Furthermore, the storage modulus $G'(\omega)$ and the loss modulus $G''(\omega)$ were calculated using the discrete Fourier transform on the long-time regime data of $G(t)$~\cite{rubinstein2003polymer},
\begin{equation}
        G'(\omega) = \omega \int_{0}^{\infty}G(t) \mbox{sin} \omega t dt 
\end{equation}
\begin{equation}
        G"(\omega) = \omega \int_{0}^{\infty}G(t) \mbox{cos} \omega t dt
\end{equation}

\begin{figure*}[bt]
        \centering
        \includegraphics[width=1.0\linewidth,bb= 0 0 1259 1067]{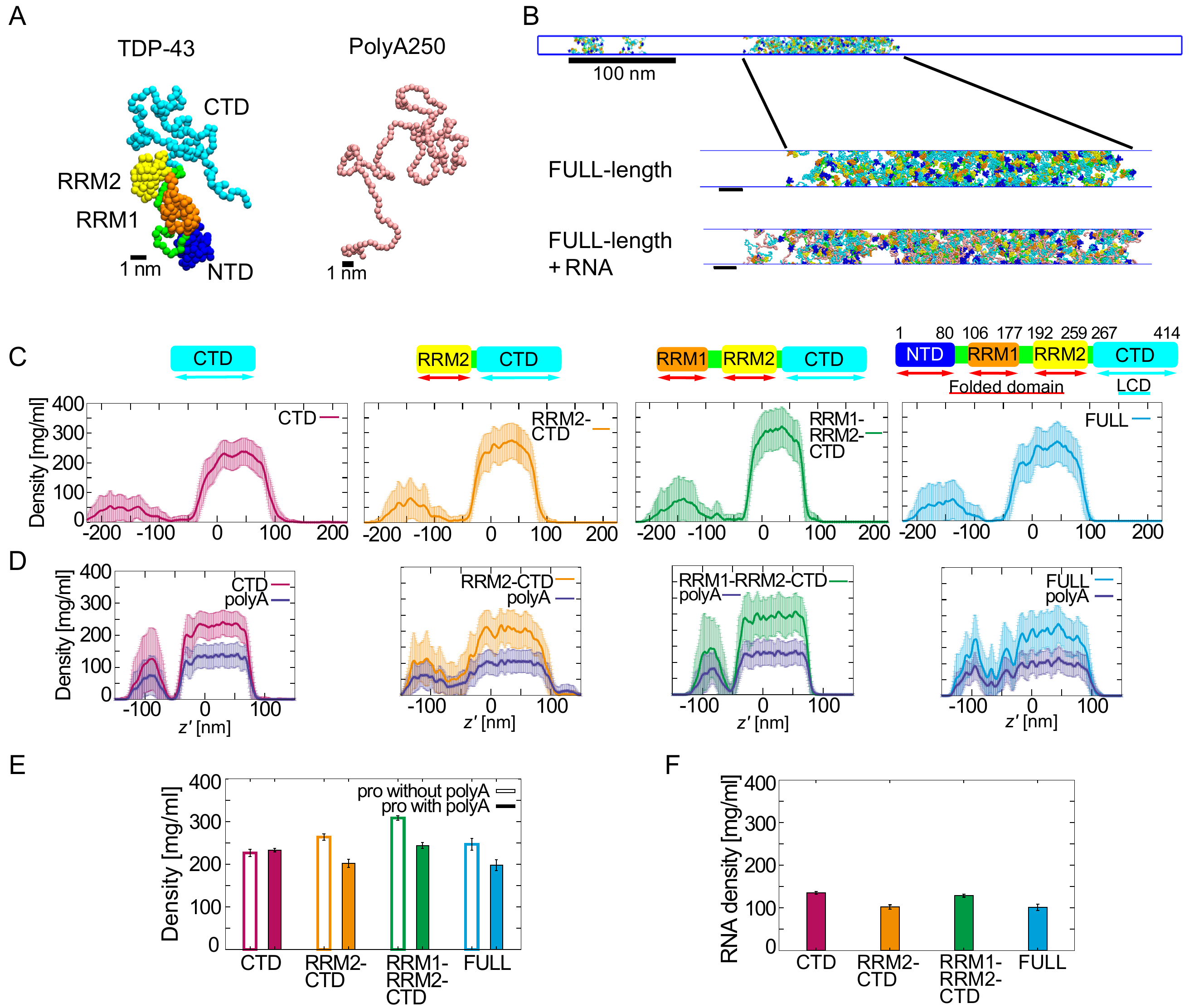}
        \caption{Phase separation of TDP-43 with different domain constructs.
        (A)~Representative structure of full-length TDP-43 and RNA (polyA250), based on the CG model. The N-terminal domain (NTD, blue), RNA recognition motif 1 (RRM1, orange), RNA recognition motif 2 (RRM2, yellow), and the C-terminal domain (CTD, cyan) are shown.
        The domain linkers are colored in green.
        (B)~Snapshots of full-length TDP-43 condensates in slab systems at $6 \, \mathrm{\mu s}$. Zoomed-in snapshots of the largest condensate is shown for full-length TDP-43 system with and without RNA. The scale bars in the zoomed-in snapshot are $10 \, \mathrm{nm}$. 
        The time-averaged density profiles of TDP-43 systems (C) without RNA and (D) with RNA. The $x$-axis represents the distance from the center of mass (COM) of the system.
        Average densities of (E) TDP-43 and (F) RNA in the dense phase.
        }
        \label{fig1}    
\end{figure*}

\begin{figure*}[tb]
        \centering
        \includegraphics[width=145 mm,bb= 0 0 984 912]{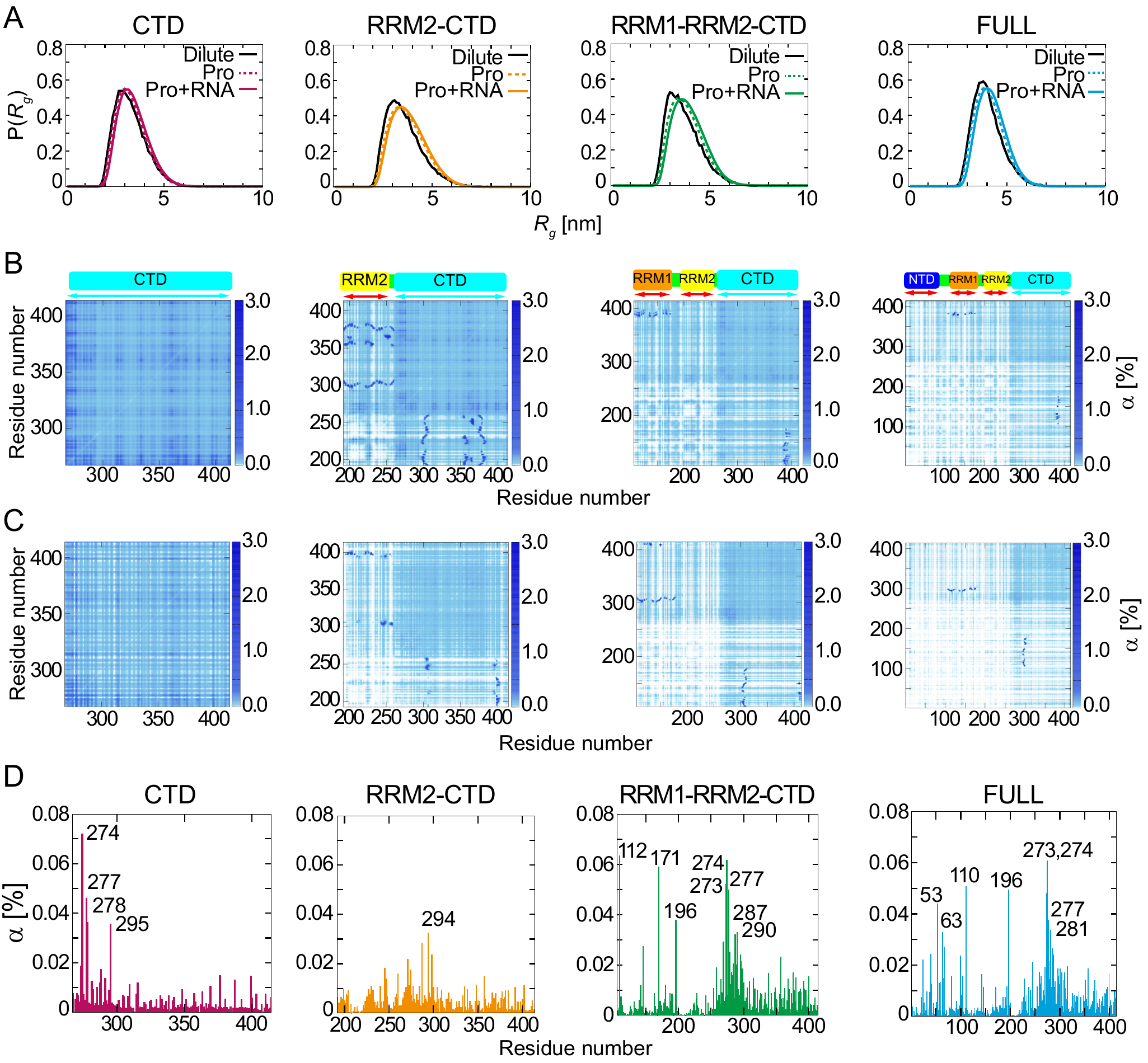}
        \caption{Molecular interactions within TDP-43 and between TDP-43 and polyA.
        (A)~Probability density functions  of the radius of gyration, P($R_{g}$).
        Dotted, dashed, and solid lines show $R_{g}$ obtained from single-chain simulations, slab simulations of TDP-43, and slab simulations of TDP-43 with polyA, respectively.
        (B)~Contact maps of intermolecular TDP-43-TDP-43 interactions in the absence and (C)~presence of polyA.
        (D)~Contact ratio of TDP-43 residues with polyA.
        Residues with a contact ratio greater than 0.15 are labeled by their residue numbers.}
        \label{fig2}
\end{figure*}

\section*{Results}
\subsection*{Multidomain Structure of TDP-43 and RNA Influence the Density of Condensates}
The phase separation of the CTD with RNA has been experimentally observed at $310 \, \mathrm{K}$~\cite{dhakal2021prion}.
To further understand how multidomain structures and RNA modulate TDP-43 phase separation, we analyzed condensates formed in slab simulations.
Previous studies suggested that the folded domains of TDP-43 preferentially assemble in the interior of condensates, while CTD regions localize at the surface~\cite{tejedor2021rna}.
In our simulations, phase separation occurred in all systems, both in the presence and absence of RNA (Figs.~\ref{fig1}C and \ref{fig1}D).
While splitting of condensates was observed, consistent with earlier work~\cite{tejedor2021rna}, we found that folded domains and CTDs were uniformly distributed within the condensates, rather than forming distinct interior or surface layers (Figs.~\ref{fig1}B-\ref{fig1}D).

We calculated the average density of TDP-43 in the largest condensates to quantify the effects of multidomain structures on phase separation~(Fig.~\ref{fig1}E).
In the absence of RNA, the densities were $226.1 \pm 8.5 \, \mathrm{mg/ml}$ for CTD, $263.3 \pm 7.3 \, \mathrm{mg/ml}$ for RRM2-CTD, $308.6 \pm 5.0 \, \mathrm{mg/ml}$ for RRM1-RRM2-CTD, and $246.7 \pm 13.5 \, \mathrm{mg/ml}$ for the full-length protein.
These results indicate that including RRM1 and RRM2 increases TDP-43 density within condensates, while adding the NTD reduces it, suggesting opposing roles of these domains in phase separation.

In the presence of polyA RNA, the densities of TDP-43 were $233.0 \pm 3.8 \, \mathrm{mg/ml}$ for CTD, $202.1 \pm 9.7 \, \mathrm{mg/ml}$ for RRM2-CTD, $243.7 \pm 6.7 \, \mathrm{mg/ml}$ for RRM1-RRM2-CTD, and $197.6 \pm 12.7 \, \mathrm{mg/ml}$ for the full-length protein (Figs.~\ref{fig1}E).
The density of CTD remained almost the same, whereas other constructs showed a decrease in protein density, suggesting that RNA preferentially interacts with RRMs and NTD.
The corresponding RNA densities were $135.5 \pm 2.9 \, \mathrm{mg/ml}$, $102.3 \pm 4.8 \, \mathrm{mg/ml}$, $128.8 \pm 3.6 \, \mathrm{mg/ml}$, and $101.4 \pm 7.3 \, \mathrm{mg/ml}$, respectively for each corresponding system (Fig.~\ref{fig1}F).
Notably, the mass ratio of TDP-43 to polyA RNA was approximately $2:1$ across all systems, indicating a consistent ratio between TDP-43 and RNA.
These results highlight the distinct contributions of TDP-43 domains to condensate formation and density, as well as the modulatory effects of RNA.

\subsection*{Multidomain and RNA-Mediated Interactions of TDP-43}
In the previous section, we demonstrated that the multidomains of TDP-43 and its interactions with RNA modulate the condensate density.
To elucidate the molecular basis underlying these effects, we analyze how TDP-43 conformations and intermolecular interactions vary under different conditions.
First, we evaluated the conformational changes of TDP-43 within condensates by comparing the radius of gyration $R_g$ of TDP-43 in the condensate with that in its isolated state in bulk solution~\cite{watanabe2024diffusion}.
As shown in Fig.~\ref{fig2}A, TDP-43 adopts an expanded conformation within the condensate.
The presence of polyA increases $R_g$ slightly, implying that both TDP-43--TDP-43 interactions and TDP-43--polyA interactions contribute to the expansion of TDP-43 conformations.

We analyzed the intermolecular interactions among TDP-43 molecules in RNA-free systems.
The contact map in Fig.~\ref{fig2}B indicates that in the CTD-only system, extensive intermolecular contacts occur, particularly among residues 280-370.
In systems incorporating folded domains, although CTD-CTD interactions remain dominant, we also observe localized interactions between the folded domains and the CTD.
For example, in the RRM2-CTD system, a cluster of CTD residues around 300, 350, and 380 interactes with RRM2 (residues 192-259), with residue 370 additionally interacting with residues around position 250 in RRM2.
In the RRM1-RRM2-CTD system, strong intramolecular interactions between RRM1 and RRM2 appear to limit additional intermolecular contacts involving RRM2, suggesting that RRM1 and RRM2 form a stable tandem configuration (Fig.~S2A).
Moreover, a region around residue 390 in the LCD forms contacts with RRM1.
The full-length system exhibits an intermolecular contact pattern similar to that of the RRM1-RRM2-CTD system.

To further understand the molecular basis of strong localized interactions at the folded domains, we categorized high-frequency residue-residue contacts (defined as interaction ration $\alpha$>1\%).
CTD-only system does not show such strong interactions (Figs.~\ref{fig2}B and S3A).
In RRM2-CTD systems, hydrophobic-hydrophobic interactions are observed, presumably occurring between hydrophobic surface patches on RRM2 and the aromatic residues of the CTD. These strong interactions are not limited to the hydrophobic patches, as other regions of RRM2 interact with GLY residues of the CTD (Figs.~\ref{fig2}B and S3A). 
RRM1-RRM2-CTD system shows similar results (Figs.~S3A).
The full system shows a higher ratio of hydrophobic interactions than other systems (Figs.~S3A).

We then compared the interaction patterns with those in RNA-containing systems (Fig.~\ref{fig2}C).
In the CTD system, overall contacts decrease in the presence of RNA.
In the CTD-RRM2 system, interactions emerge between the CTD (around residue 395) and RRM1, indicating that RNA alters the CTD's interaction sites.
Similarly, in the CTD-RRM2-RRM1 system, RRM2-CTD interactions appear at CTD sites around residues 310 and 410, distinct from the RNA-free condition.
In the full-length system, interaction at CTD residue 290 is different from the RNA-free system.
Categorization of interaction types reveals that in the CTD-RRM2, CTD-RRM2-RRM1 and full-length systems, hydrophobic interactions in the RNA-containing systems were reduced in RNA-containing conditions (Fig.~S3B).
We speculate that this reduction is due to the strong short-ranged interactions, as indicated by a high values of the stickiness parameter of the adenine bead, replacing some hydrophobic protein-protein interactions.

Finally, we analyzed TDP-43-polyA contacts (Fig.~\ref{fig2}D).
In the CTD system, residues that frequently contact polyA include Gly residues in the 270-295 region (specifically G274, G277, G278, and G295).
In the RRM2-CTD system, residue R294 exhibits frequent contacts.
In the RRM1-RRM2-CTD system, polyA interacts with residues in both RRM1 (P112, R171, and G196) and CTD (S273, G274, G277, G287, and G290).
In the full-length system, interactions occur at the NTD (G53, A63), RRM1 (G110), RRM2 (G196), and CTD (S273, G274, G277, and G281).
Consistently, our analysis of interaction types shows that the TDP-43--polyA interactions occur predominantly at polar residues, specifically glycines, and, in the RRM2-CTD and full-length systems, some hydrophobic interactions are also observed (Figs.~\ref{fig2}D and S3C).
These results indicate strong interactions between polyA and both the CTD and the folded domains of TDP-43.

Our results reveal that in all systems the CTD engages in multivalent interactions, forming multiple, broadly distributed contacts.
In contrast, the folded domains tend to form more localized and specific interactions with distinct regions of the CTD.
These specific interaction patterns are influenced by the presence of other folded domains and RNA.
For example, the interaction between RRM2 and CTD is diminished by the presence of RRM1, consistent with a stable tandem arrangement between these two domains.
Furthermore, the addition of polyA alters the interaction landscape of TDP-43.

\begin{figure*}[tb]
        \centering
        \includegraphics[width=0.7\linewidth,bb= 0 0 904 1219]{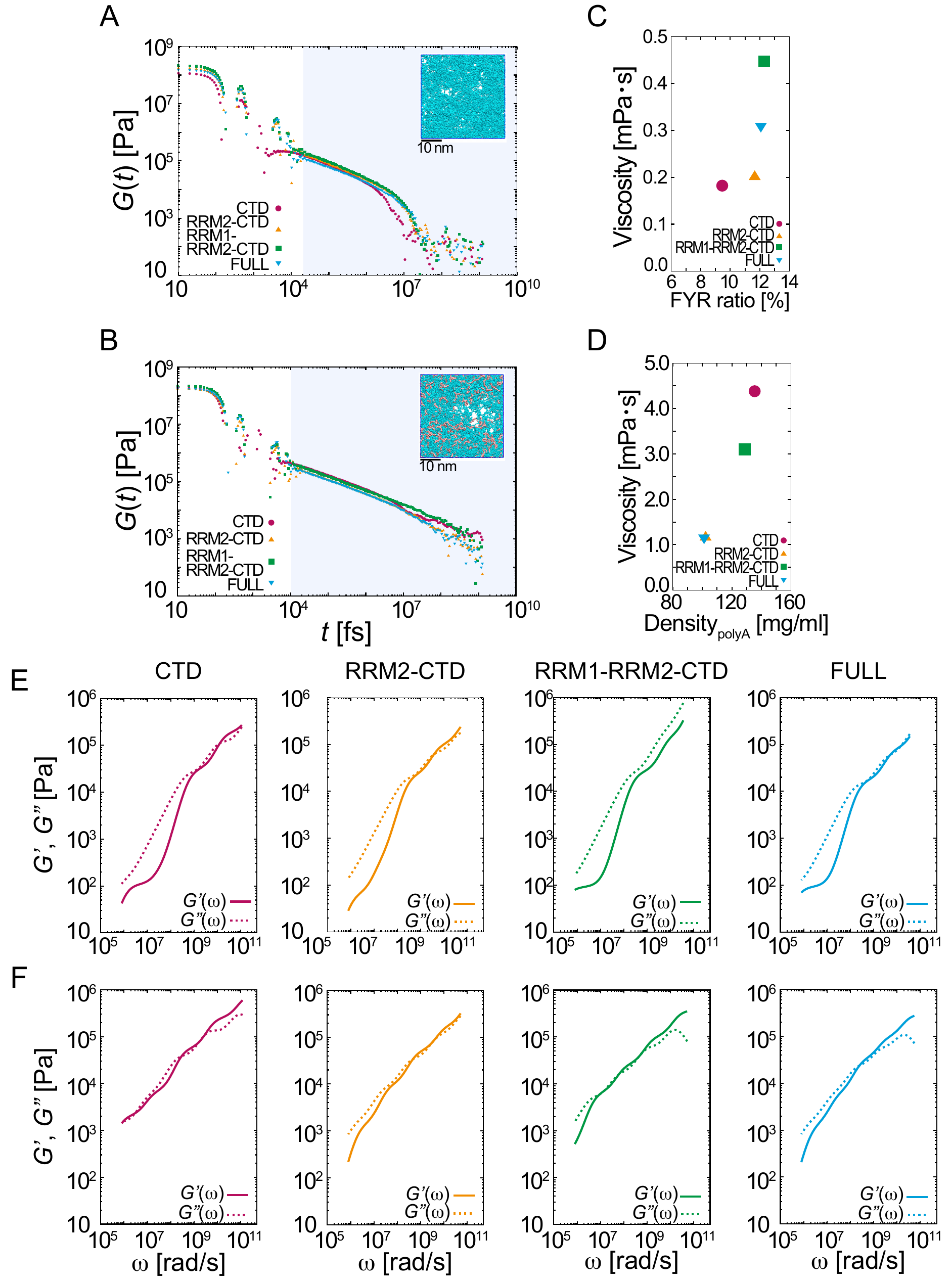}
        \caption{Viscoelasticity of condensates.
        (A)~Shear modulus of TDP-43 condensates without polyA and (B)~with polyA.
        The inset shows a snapshot of the simulation system at $15 \, \mathrm{\mu s}$.
        (C)~Correlation between viscosity and the ratio of sticker residues (F, Y, and R) in the protein in TDP-43 systems without polyA.
        (D)~Correlation between viscosity and the polyA density in the simulations of TDP-43 with polyA.
        (E)~ Elastic modulus ($G'$, solid line) and loss modulus ($G''$, dashed line) in TDP-43 systems without polyA.
        (F)~The same quantities as in (E) for TDP-43 systems with polyA.}
        \label{fig3}
\end{figure*}

\begin{figure*}[tb]
    \centering
    \includegraphics[width=0.8\linewidth,bb= 0 0 1033 568]{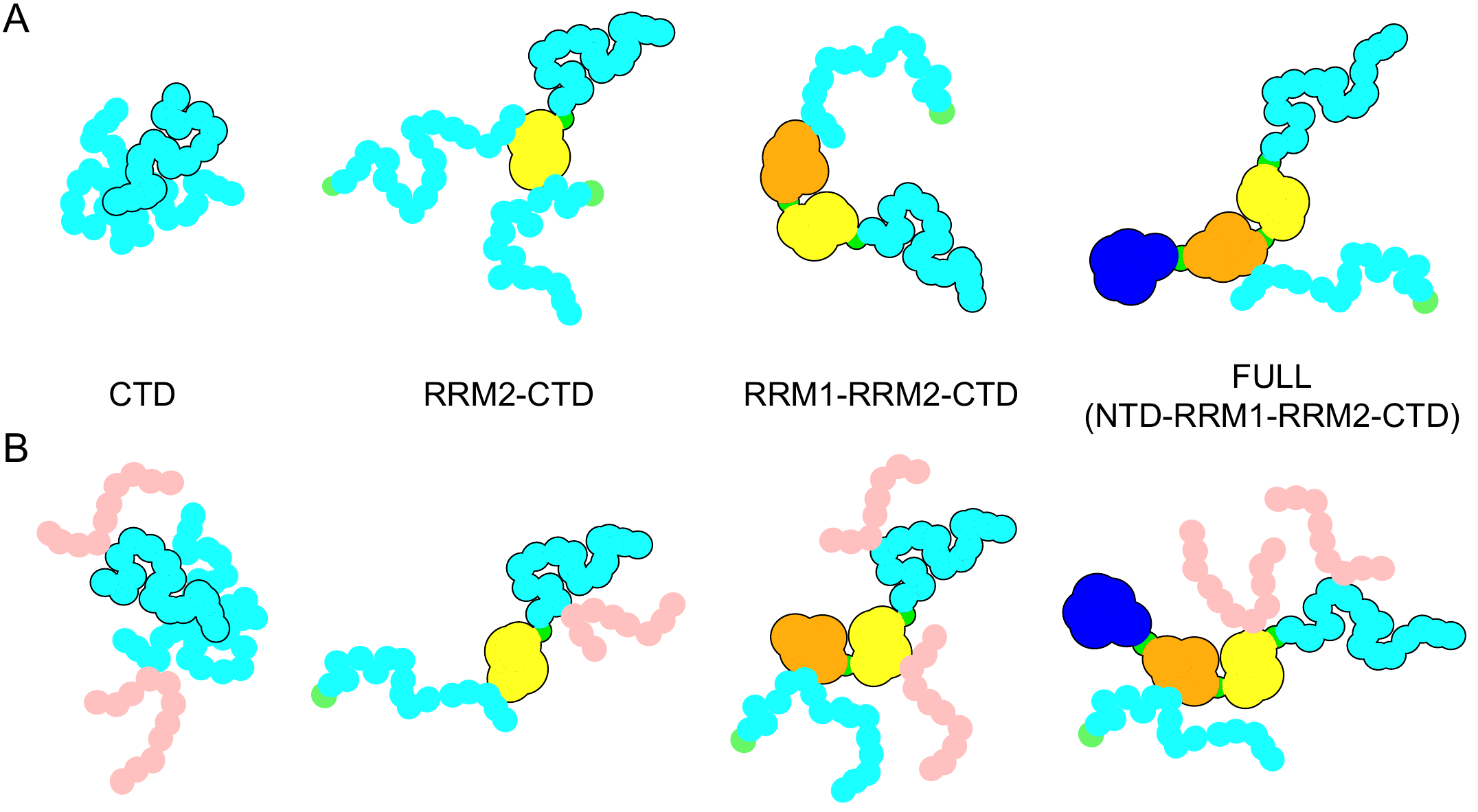}
    \caption{Schematic of molecular interaction rearrangements depending on the domain construct and the presence of RNA in (A)~TDP-43 alone systems and (B)~TDP-43 with polyA systems.
    The CTD domain (cyan), RRM2 domain (yellow), RRM1 domain (orange), NTD domain (blue), linkers (green), and polyA (pink) are shown.}
    \label{fig4}
\end{figure*}

\subsection*{Multidomains of Proteins and RNA Modulate Viscoelasticity of Condensates}
To understand how multidomain protein structures and RNA modulate viscoelasticity, we first calculated the shear modulus $G(t)$ from additional simulations conducted in cubic systems (see Figs.~\ref{fig3} and S4, details in Methods).
According to the Maxwell model (Eq.~\ref{eq_maxwell}), $G(t)$ can be expressed as a sum of multiple exponential relaxation modes.
In the time region beyond $10 \, \mathrm{ps}$, a power-law decay is observed (Figs.~\ref{fig3}A and B), indicating the presence of multiple relaxation modes over a broad timescale.
In the CTD-only system, $G(t)$ declined significantly after $2 \, \mathrm{ns}$, suggesting that this corresponds to the longest relaxation time (Fig.~\ref{fig3}A).
In contrast, systems containing folded domains exhibited the longest relaxation time of approximately $10 \, \mathrm{ns}$.
The faster relaxation of the shear modulus in the CTD-only system, compared to systems with folded domains, is due to weaker intermolecular interactions and lower network connectivity.
Indeed, the presence of polyA, which enhances network connectivity among IDPs, further extends the relaxation time, even in the CTD-polyA system (Fig.~\ref{fig3}B).
The CTD and RRM1-RRM2-CTD systems exhibited higher $G(t)$ values than the RRM2-CTD and FULL systems.
Interestingly, in the presence of polyA chains, neither the presence nor the number of folded domains directly affected $G(t)$.

We then calculated the viscosity $\eta$ by integrating $G(t)$ (Eq.~\ref{eq_viscosity}).
In TDP-43 alone systems, the viscosity followed the order: CTD ($1.82 \times 10^{-4} \, \mathrm{Pa \cdot s}$) $<$ RRM2-CTD ($2.03 \times 10^{-4} \, \mathrm{Pa \cdot s}$) $<$ FULL ($3.08 \times 10^{-4} \, \mathrm{Pa \cdot s}$) $<$ RRM1-RRM2-CTD ($4.47 \times 10^{-4} \, \mathrm{Pa \cdot s}$).
This order changed when polyA chains were added: FULL ($1.14 \times 10^{-3} \, \mathrm{Pa \cdot s}$) $<$ RRM2-CTD($1.18 \times 10^{-3} \, \mathrm{Pa \cdot s}$) $<$ RRM1-RRM2-CTD ($3.10 \times 10^{-3} \, \mathrm{Pa \cdot s}$) $<$ CTD ($4.38 \times 10^{-3} \, \mathrm{Pa \cdot s}$).
Moreover, we examined the correlations between viscosity and other physical properties.
Previous studies have suggested a positive correlation between the ratio of sticker residues (F, Y and R) within sequences and viscosity among various LCDs~\cite{tejedor2023time}.
We found that this relationship held in TDP-43 alone systems, indicating that sticker interactions play a crucial role in TDP-43 alone systems (Figs.~\ref{fig3}C and S5).
However, when polyA chains were added, this correlation was disrupted, and instead, the polyA density within the dense phase showed a strong correlation with viscosity (Figs.~\ref{fig3}D and S6).
In the physical model of monopolymer systems (Rouse chain model), viscosity is proportional to molecular mass~\cite{tejedor2021rna}.
However, this trend was not observed in our multidomain systems (Figs.~S5 and S6).
This suggests that folded structures and heterogeneous interactions in our systems give rise to more complex behavior than that predicted by simple polymer models.

Finally, we analyzed the frequency response by calculating the storage modulus $G'$ and the loss modulus $G''$. 
Generally, condensates behave as elastic materials at high frequencies ($G'(\omega) > G''(\omega)$) and as liquid-like materials at low frequencies ($G'(\omega) < G''(\omega)$)~\cite{jawerth2020protein,alshareedah2021, alshareedah2024}.
The frequency at which $G'(\omega) = G''(\omega)$ is the crossover frequency where the material behaves equally as an elastic solid and a fluid.
In the CTD-only system, the crossover frequency was around $\omega = 10^{9} \, \mathrm{rad/s}$, with similar values observed in the RRM2-CTD and FULL systems.
In contrast, the RRM1-RRM2-CTD system showed lower $G'(\omega)$ values, with viscosity dominating across all frequencies.
The addition of polyA to the CTD system increased $G'(\omega)$, resulting in a frequency range of $10^{6}-10^{9} \, \mathrm{rad/s}$ where elastic and viscous contributions were comparable.
A similar increase in elasticity was observed in the other systems (RRM2-CTD, RRM1-RRM2-CTD, and FULL).
The crossover frequency was approximately $<10^{9} \, \mathrm{rad/s}$ in the CTD system, $>10^{8} \, \mathrm{rad/s}$ in the RRM2-CTD system, $10^{7}-10^{9} \, \mathrm{rad/s}$ in the RRM1-RRM2-CTD system, and $2 \times 10^{8} \, \mathrm{rad/s}$ in the FULL system.
These results suggest that RNA plays a more dominant role in modulating the viscoelasticity of condensates than variations in molecular interactions within TDP-43.

\section*{Discussion}
The phase separation of TDP-43 is thought to have multiple triggers, including the IDRs of the CTD, RRMs, and NTD, while RNA is believed to maintain the fluidity of TDP-43 condensates, preventing fibril formation~\cite{maharana2018rna,carter2021n,tziortzouda2021triad,lang2024tdp}.
TDP-43 remains liquid-like state in the nucleus due to the RNA-rich conditions~\cite{maharana2018rna}, while inclusion into cytoplasm occur in conjugate with mRNA as stress responses and exposure to cytoplasm is an initial step of liquid-to-solid transition~\cite{tziortzouda2021triad,lang2024tdp}.    
However, the roles of individual domains and their cooperative molecular interactions within TDP-43 condensates, particularly in the presence of multiple domains and RNA chains, have remained unclear.
In this study, we performed CG MD simulations of TDP-43 condensates using four types of domain constructs, both with and without polyA.
We observed that intra- and intermolecular interaction patterns within the condensates changed cooperatively among the IDRs, folded domains, and polyA chains (Fig.~\ref{fig4}).
Furthermore, our results demonstrated that RNA significantly influences the viscoelasticity of the condensates.

Our simulations of TDP-43 and TDP-43-RNA mixing systems partially capture the cytoplasmic and nuclear conditions, respectively, and provide molecular insights into TDP-43 condensates in these environments. 
Simulations of TDP-43 alone suggest that while CTD interactions primarily drive condensation, relatively strong interactions between RRM1 and the CTD enhance molecular interaction network within condensates. These interactions may alter the conformational stability of both the RRMs and the CTD, leading to misfolding of RRMs in cytoplasmic stress granules \cite{wang2013truncated,yan2024intra} and promoting the formation of secondary structures in the CTD, including transient helices and amyloid $\beta$-sheets \cite{conicella2016mutations,conicella2020tdp}.
On the other hand, in RNA-containing systems, our results show that RNA is incorporated into TDP-43 condensates through sticky interactions of RNA mainly with Gly residues of TDP-43, rather than via ionic interactions.
Consequently, TDP-43--RNA interactions replace the TDP-43 intermolecular interactions observed in RNA-free systems (Fig.~\ref{fig4}). 
This replacement mechanism may be crucial to prevent amyloid formation and to maintain the fluidity of TDP-43 condensates in the nucleus. Indeed, our results show that the inclusion of RNA increases the elasticity of the condensates compared to those formed solely by TDP-43, suggesting that RNA acts as a cross-linker, even at a relatively short length of 250 nucleotides, and stabilizes the network in condensates~\cite{sanchez2022rna}.

While CG models provide valuable insights, their accuracy is inherently limited by the simplifications made to reduce computational cost.
First, residue-level CG models may not fully capture specific molecular interactions, particularly those involving side-chain interactions and backbone conformational flexibility.
For instance, particular caution is required when analyzing oligomerization at the NTD and transient helix formation in the CTD, as these processes may involve intricate side-chain interactions and backbone flexibility that are not well represented in CG models.
Second, implicit solvent models assume low viscosity, leading to artificially accelerated dynamics~\cite{watanabe2024diffusion}.
In our calculations, the estimated viscosity is approximately 3-4 orders of magnitude lower than experimental values~\cite{gopal2017amyotrophic}.
By enabling large-scale simulations that are computationally infeasible with AT models, CG MD simulations provide valuable insights into the phase behavior, structural organization, and dynamic properties of multidomain proteins, thereby advancing the molecular-level understanding of biophysical properties underlying biological phenomena.

\section*{Acknowledgments}
This work was supported by JSPS KAKENHI Grant Number 21H05728 and JST PRESTO Grant Number JPMJPR22EE, Japan.
We thank computer FUGAKU (ID: hp230327, hp240302) for providing computing resources for this work.
I.Y. acknowledge supports from JSPS KAKENHI Grant Number JP23KJ1918) and JST ACT-X Grant Number JPMJAX24LJ.


%

\clearpage

\onecolumngrid
{
    \center \bf \large 
    Supplementary Materials\vspace*{1cm}\\ 
    \vspace*{0.0cm}
}

\setcounter{figure}{0}

\begin{figure*}[hbt!]
        \centering
        \includegraphics[width=0.9\linewidth,bb= 0 0 474 326]{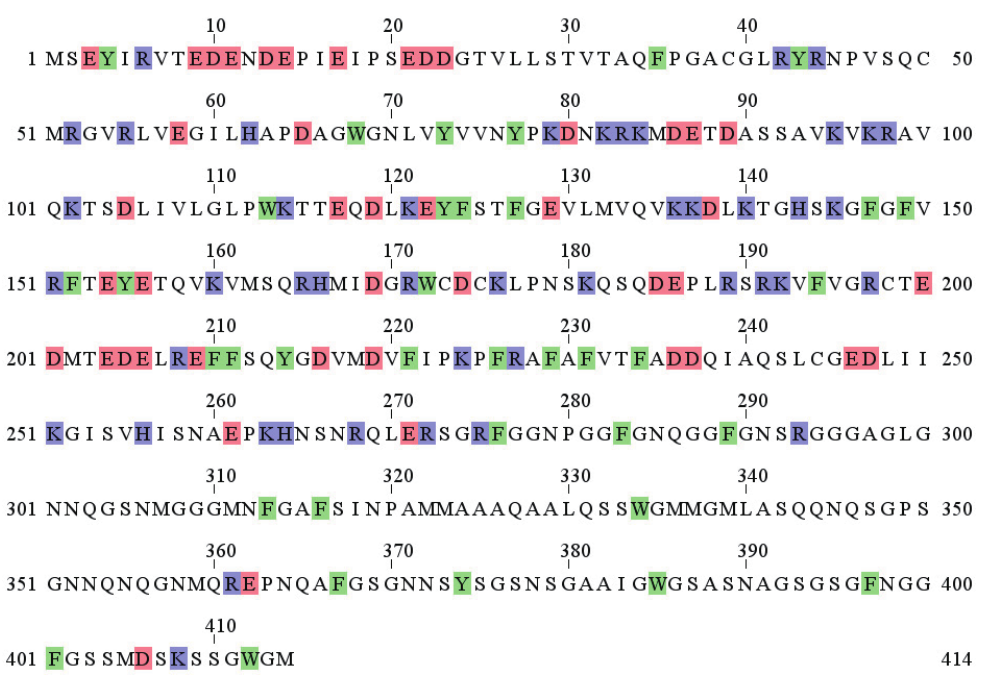}
        \caption{Sequence of full-length TDP-43.
        Positively charged, negatively charged, and aromatic amino acids are marked in blue, red, and green, respectively.}
        \label{figsup_aa}
\end{figure*}

\begin{figure*}[hbt!]
    \centering
    \includegraphics[width=0.9\linewidth,bb= 0 0 926 401]{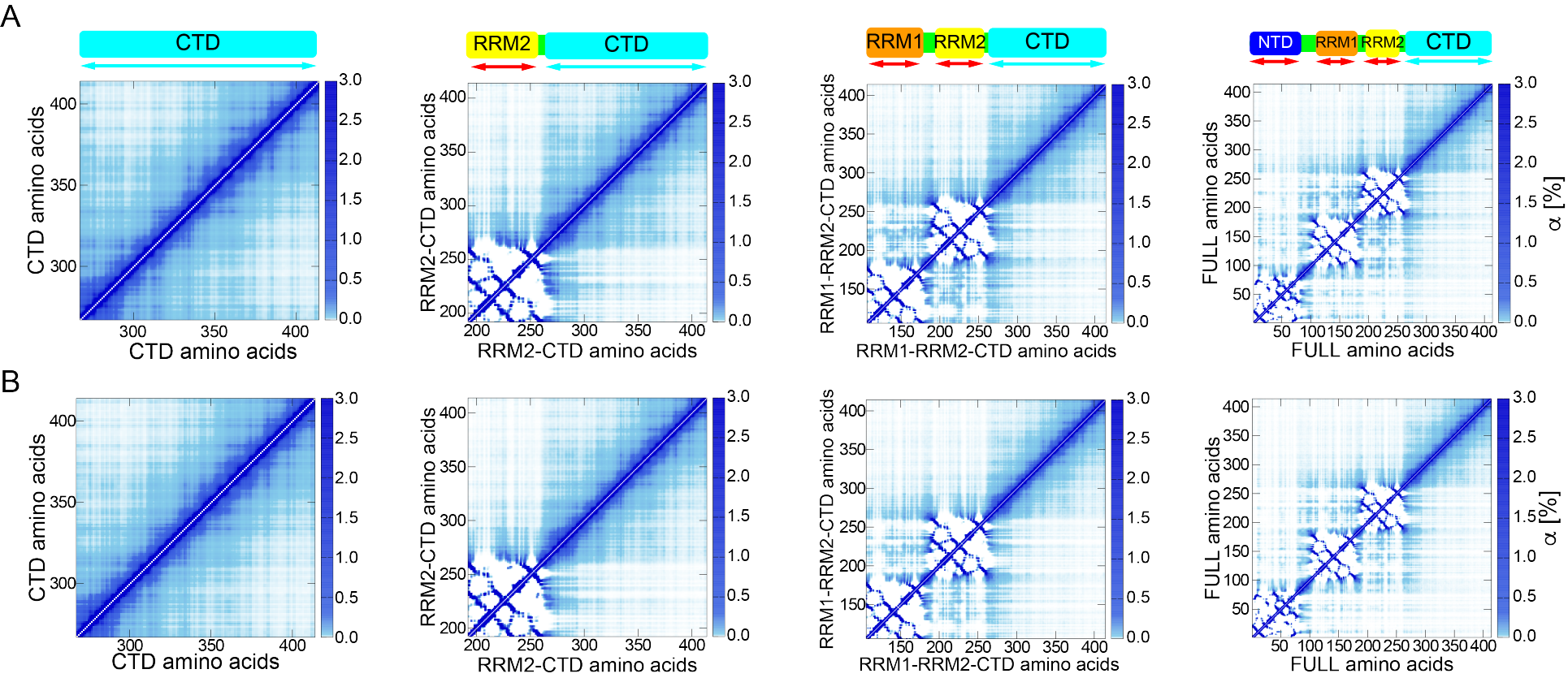}
    \caption{Intramolecular interactions of TDP-43 (A)~in the absence of polyA and (B)~in the presence of polyA.}    
    \label{fig_sup_conmap_intra}
\end{figure*}

\begin{figure*}[hbt!]
    \centering
    \includegraphics[width=0.9\linewidth,bb= 0 0 1153 669]{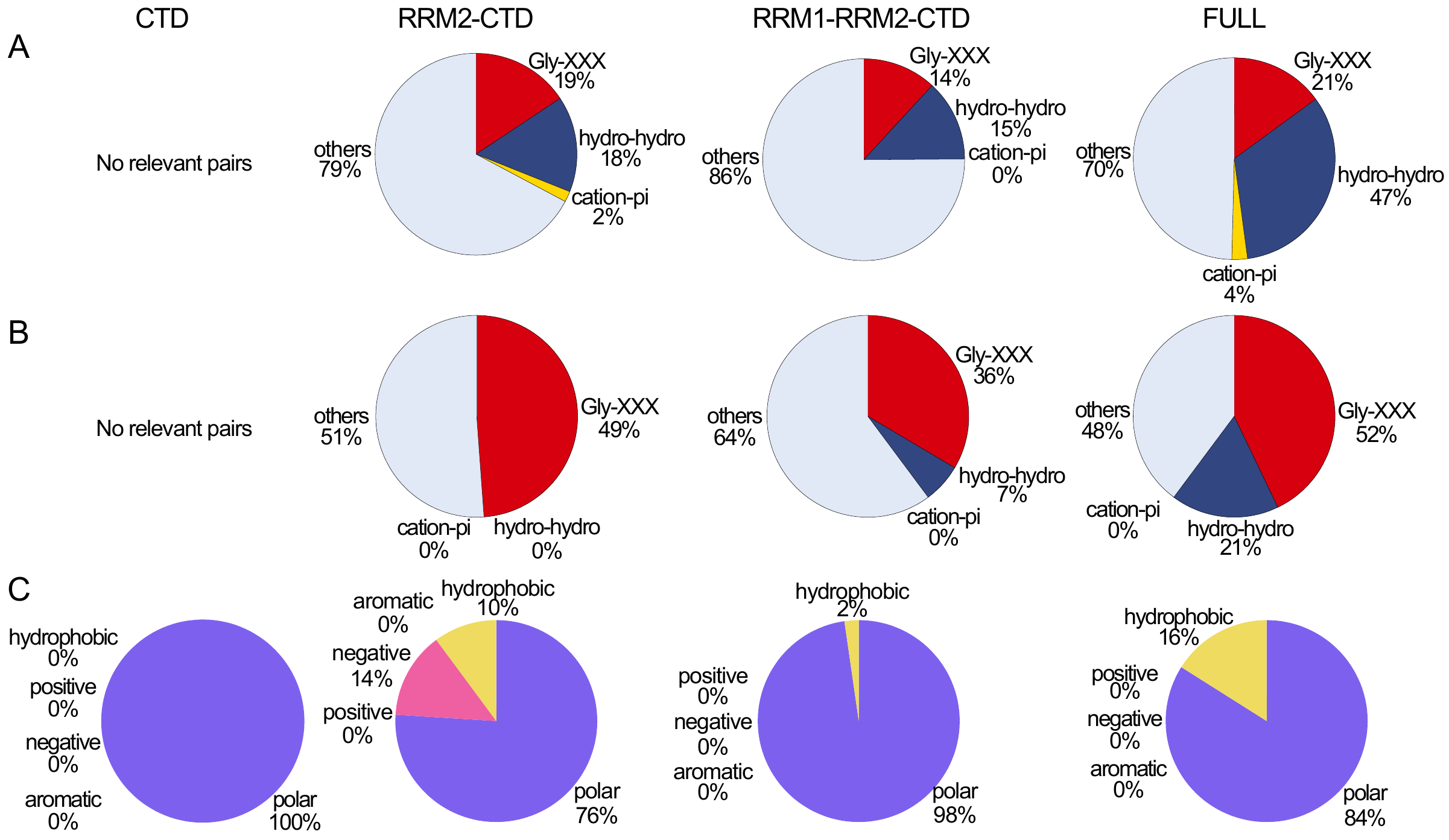}
    \caption{Classification of intermolecular interactions.
    (A, B)~In the TDP-43--TDP-43 interactions, residue pairs with contact ratio higher than 1.0\% are classified into Gly--other (XXX), hydrophobic--hydrophobic (HB--HB), cation--$\pi$, and other types of interactions.
    (A)~The classicification in the TDP-43 alone systems and (B)~TDP-43 with polyA systems are shown. 
    (C)~In the TDP-43 with polyA systems, TDP-43 residues that interacted with polyA at high frequency ($\alpha$>0.015\%) were classified.}
    \label{fig_sup_conmap_local}
\end{figure*}

\begin{figure*}[hbt!]
        \centering
        \includegraphics[width=0.9\linewidth,bb= 0 0 1042 1115]{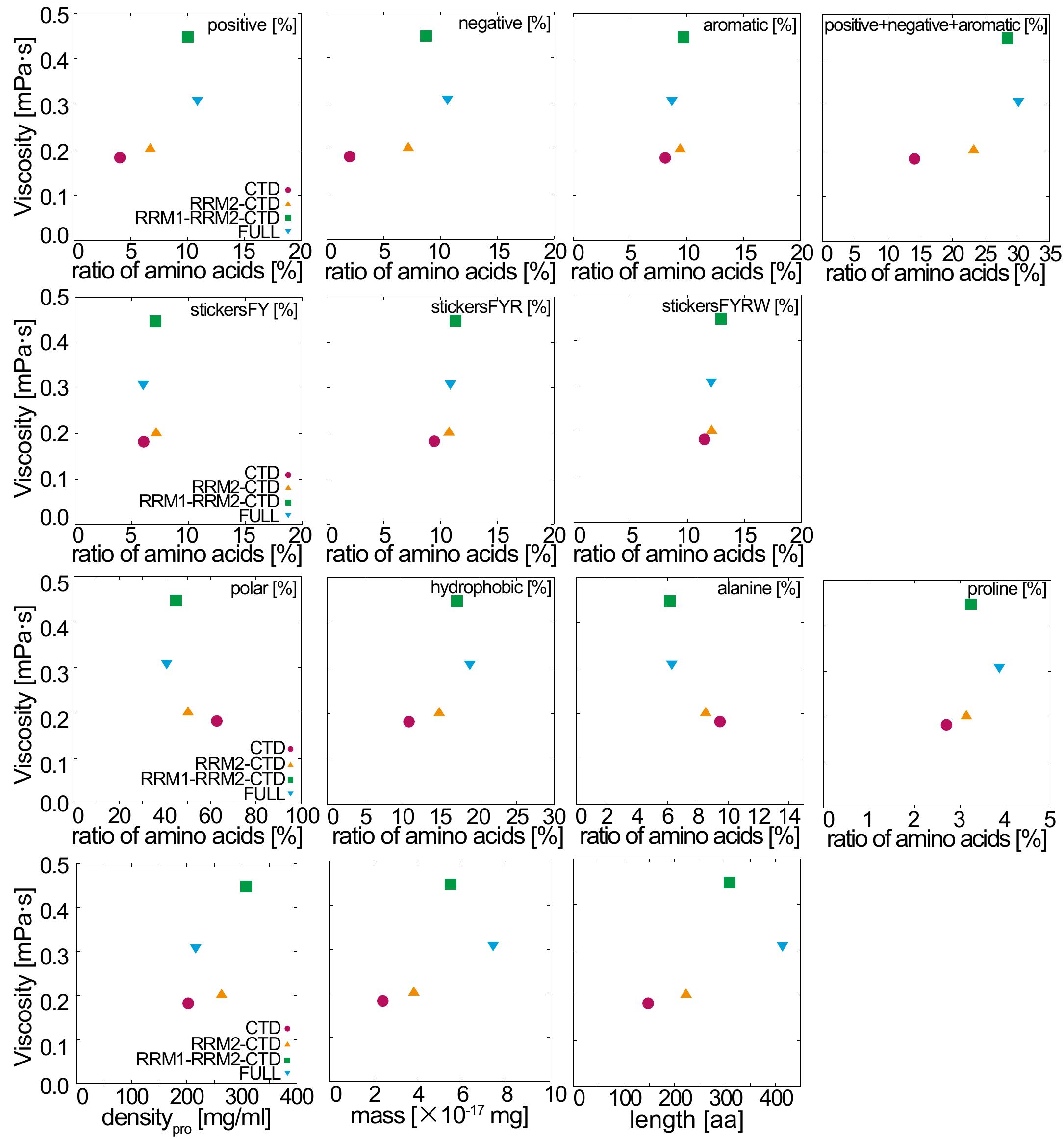}
        \caption{Correlation between condensate viscosity in the absence of polyA and various molecular properties.
        The viscosity is plotted against the number of positively charged residues (R, H, and K) (1st row, 1st column), negatively charged residues (D and E) (1st row, 2nd column), aromatic residues (F, Y, and W) (1st row, 3rd column), and the total number of positively charged, negatively charged, and aromatic residues (1st row, 4th column).
        It is also plotted against the number of stickers (F and Y) (2nd row, 1st column), stickers (F, Y, and R) (2nd row, 2nd column), and stickers (F, Y, R, and W) (2nd row, 3rd column), as well as the number of polar residues (3rd row, 1st column), hydrophobic residues (3rd row, 2nd column), alanine residues (3rd row, 3rd column), and proline residues (3rd row, 4th column).
        Additionally, the correlation with protein density (4th row, 1st column), molecular weight (4th row, 2nd column), and chain length (4th row, 3rd column) is shown.}
        \label{figsup_eta_nA}
\end{figure*}

\begin{figure*}[hbt!]
        \centering
        \includegraphics[width=1.0\linewidth,bb= 0 0 1287 1115]{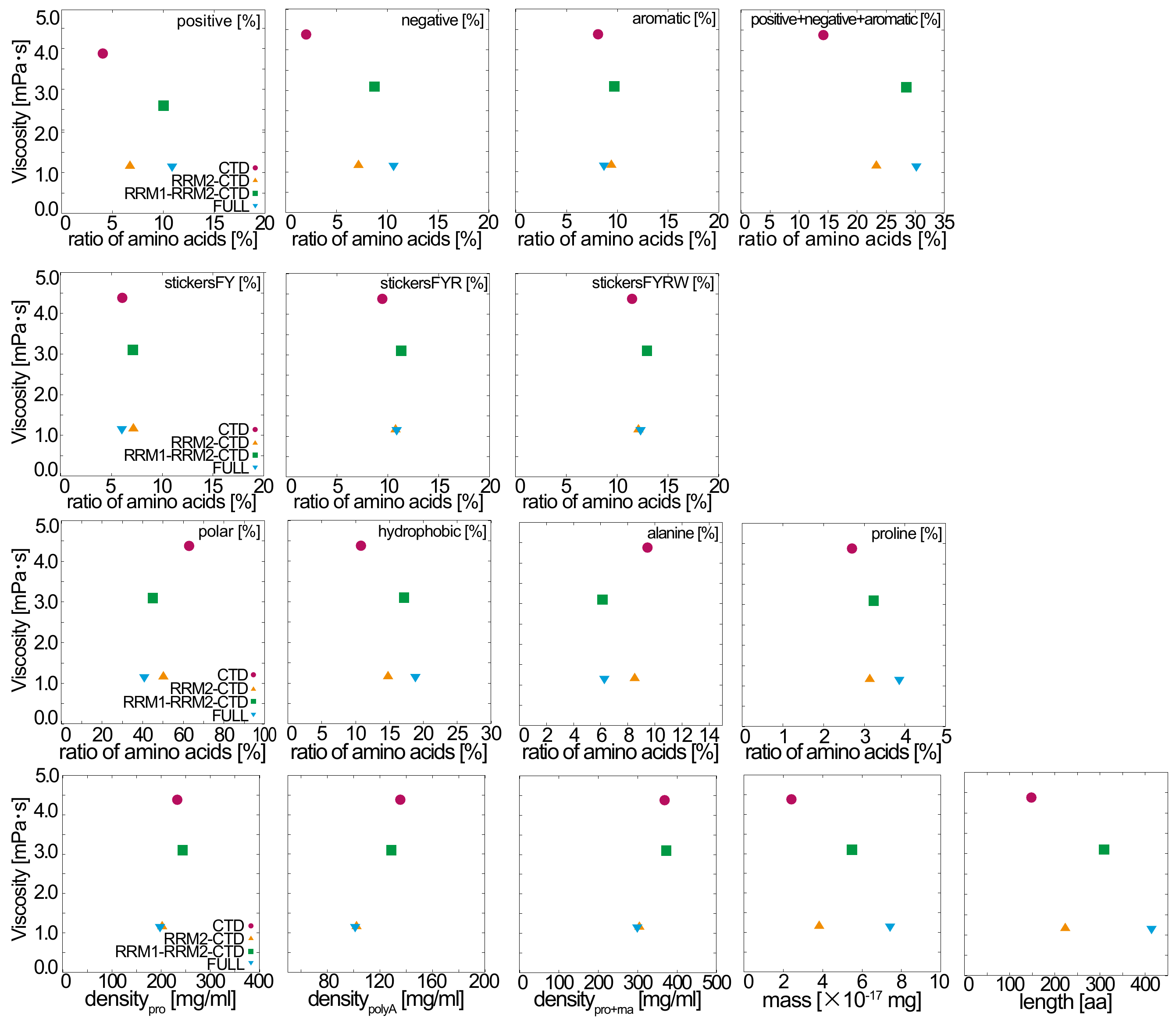}
        \caption{Correlation between condensate viscosity in the presence of polyA and various molecular properties.
        The viscosity is plotted against the number of positively charged residues (R, H, and K) (1st row, 1st column), negatively charged residues (D and E) (1st row, 2nd column), aromatic residues (F, Y, and W) (1st row, 3rd column), and the total number of positively charged, negatively charged, and aromatic residues (1st row, 4th column).
        It is also plotted against the number of stickers (F and Y) (2nd row, 1st column), stickers (F, Y, and R) (2nd row, 2nd column), and stickers (F, Y, R, and W) (2nd row, 3rd column), as well as the number of polar residues (3rd row, 1st column), hydrophobic residues (3rd row, 2nd column), alanine residues (3rd row, 3rd column), and proline residues (3rd row, 4th column).
        Additionally, the correlation with protein density (4th row, 1st column), molecular weight (4th row, 2nd column), and chain length (4th row, 3rd column) is shown.}
        \label{figsup_eta_wA}
\end{figure*}

\end{document}